 \documentclass[final,3p,times]{elsarticle}
\usepackage{graphicx} 
\usepackage{tabularx}
\usepackage{amsmath}
\usepackage{amssymb}
\usepackage{subfigure}
\usepackage{subfigure}
\usepackage{comment}
\usepackage{color}

\usepackage{amssymb,amsmath,bm}
\definecolor{lightgray}{gray}{0.75}

\usepackage{fancyhdr}
\usepackage{lipsum}
\usepackage[dvipsnames]{xcolor}

\usepackage{hyperref}

\usepackage{cleveref}

\newcommand\myshade{85}
\colorlet{mylinkcolor}{violet}
\colorlet{mycitecolor}{YellowOrange}
\colorlet{myurlcolor}{Aquamarine}

\hypersetup{
  colorlinks=true,
  linkcolor  = mylinkcolor!\myshade!black,
  citecolor  = mycitecolor!\myshade!black,
  urlcolor   = myurlcolor!\myshade!black
}
\journal{CMA report}

\begin{document}

\begin{frontmatter}

\title{ \LARGE Development of a unified high-order nonhydrostatic multi-moment constrained finite volume dynamical core: derivation of  flux-form governing equations in the general curvilinear coordinate system}

\author[cnwp]{Xingliang Li} 
\author[xjtu]{Chungang Chen \corref{cor}}
\author[cnwp]{Xueshun Shen}
\author[titech]{Feng Xiao}

\address[cnwp]{Center of Numerical Weather Prediction of NMC, China Meteorological Administration,  46 Zhongguancun South St., Beijing 100081, China }
\address[xjtu]{State Key Laboratory for Strength and Vibration of Mechanical Structures \& School of Human Settlement and Civil Engineering, Xi'an Jiaotong University, 28 Xianning West Road, Xifan, Shaanxi, 710049, China}
\address[titech]{ Department of Mechanical Engineering, Tokyo Institute of Technology, Tokyo 226-8502, Japan }

\cortext[cor]{Corresponding Address: Xi'an Jiaotong University, 28 Xianning West Road, Xifan, Shaanxi, 710049, China. Email address: cgchen@xjtu.edu.cn}
%\cortext[cor]{Corresponding Address: Center of Numerical Weather Prediction, China Meteorological Administration,  46 Zhongguancun South St., Beijing 100081, China. Email address: lixliang@cma.gov.cn}

\begin{abstract}
In the manuscript we have derived the flux-form atmospheric governing equations in the general curvilinear coordinate system which is used by a high-order nonhydrostatic multi-moment constrained finite volume (MCV) dynamical core, and given the explicit formulations in the shallow-atmosphere approximation. In general curvilinear coordinate $x^i(i=1,2,3)$, unlike the Cartesian coordinate, the base vectors are not constants either in magnitude or direction. Following the representations such as base vectors, vector and tensor and so on in general curvilinear coordinate, we can obtain the differential relations of base vectors, the gradient and divergence operator etc. which are the component parts of the atmospheric governing equation. Then we apply them in the two specific curvilinear coordinate system: the spherical polar and cubed-sphere coordinates that are adopted in high-order nonhydrostatic MCV dynamical core. By switching the geometrics such as the metric tensors (covariant and contravariant), Jacobian of the transformation, the Christoffel symbol of the second kind between the spherical polar and cubed-sphere coordinates, the resulting flux-form governing equations in the specific coordinate system can be easily achieved. Of course, the Cartesian coordinate can be recovered. Noted that the projection metric tensors like spherical polar system and Cartesian coordinate become simple due to orthogonal properties of coordinate.
\end{abstract}


\end{frontmatter}

%% create the table of contents
%\clearpage
%\tableofcontents
\clearpage

\section{3D compressible non-hydrostatic Euler equation set}
An inviscid, no-heat conducting fluid in 3-dimensional motion is
governed by the local conservation of mass density $\rho$, momentum
density $\rho\mathbf{u}$ and potential temperature $\theta$. The
geometric form of these equation on the rotating Earth (angular velocity $\mathbf{\Omega}$), independent of any coordinate
basis, can be written as
\begin{align}
\frac{\partial}{\partial t}\begin{pmatrix}
     \rho\\
     \rho \mathbf{u} \\
     \rho \theta
   \end{pmatrix}+\nabla \cdot
\begin{pmatrix}
     \rho \mathbf{u}\\
     \overleftrightarrow{\mathbf{T}} \\
     \rho \theta
   \end{pmatrix}=
\begin{pmatrix}
     0\\
     -\rho \nabla \Phi -2\mathbf{\Omega} \times \rho \mathbf{u}  \\
     0
   \end{pmatrix} \label{eq:vectorformEq}
\end{align}
where the momentum tensor is
\begin{align}\label{eq:momen_tensor}
\overleftrightarrow{\mathbf{T}} =\rho \mathbf{u} \otimes \mathbf{u}
+ \overleftrightarrow{\mathbf{G}}p
\end{align}
$\overleftrightarrow{\mathbf{G}}$ is metric tensor due to coordinate
transformation, and potential $\Phi$ satisfies the Poisson equation
$\Delta \Phi=4\pi g_c \rho$, and $g_c$ is the universal gravitational
constant.  

More specically the momentum tensor could be rewritten as
\begin{align}\label{eq:momen_tensor_components}
\overleftrightarrow{\mathbf{T}} &={T}^{ij}\mathbf{a}_i\mathbf{a}_j \\
  {T}^{ij} & =\rho u^iu^j+G^{ij}p
\end{align}
where $\mathbf{a}_i$ and $\mathbf{a}_j$ are the covariant base vectors, ${T}^{ij} $ is the contravariant components of $\overleftrightarrow{\mathbf{T}}$, $G^{ij}$ is the contravariant metric in the curvilinear coordinates and the indices $i$ and $j$ span by $(1,2,3) \text{ or } (\xi,\eta,r) $ of the coordinate line $(x^1,x^2,x^3)$.

\section{The nonhydrostatic governing equations in general curvilinear form}
Based on the representations in appendix \ref{label:appendixA}, we now can express the governing equations \eqref{eq:vectorformEq} in general curvilinear form as
\begin{align} % requires amsmath; align* for no eq. number
   \frac{\partial \rho}{\partial t} + \frac{1}{\sqrt{G}} \left[ \frac{\partial (\sqrt{G}\rho u^j)}{\partial x^j} \right] &=0, \label{eq:curvicoord1} \\
   \frac{\partial \rho u^i}{\partial t} +   \frac{1}{\sqrt{G}}\frac{\partial }{\partial x^j} \left[\sqrt{G}(\rho u^i u^j +G^{ij}p) \right] +\Gamma_{jk}^i(\rho u^ju^k+G^{jk}p) &= F_C^i-\rho g G^{3i}, \label{eq:curvicoord2}\\
   \frac{\partial \rho \theta}{\partial t} + \frac{1}{\sqrt{G}} \left[ \frac{\partial (\sqrt{G}\rho\theta u^j)}{\partial x^j} \right] &=0,  \label{eq:curvicoord3} \\
     \frac{\partial \rho q_k}{\partial t} + \frac{1}{\sqrt{G}} \left[ \frac{\partial (\sqrt{G}\rho q_k u^j)}{\partial x^j} \right] &=0, \label{eq:curvicoord4}
\end{align}
where $u^i$ is contravariant velocity in curvilinear coordinates, $G_{ij}$ being fundamental  metric tensor, $\sqrt{G}=\det(G_{ij})^{1/2}$ is the Jacobian of the transformation, $G^{ij}=G_{ij}^{-1}$, and $i,j,k \in (1,2,3) \textrm{ or } (\xi,\eta,r)$, $q_k$ is the moisture species. The Christoffel symbol of the second kind $\Gamma_{jk}^i$, namely contravariant derivative of the covariant basis, is
\begin{align} % requires amsmath; align* for no eq. number
   \Gamma_{jk}^i = \frac{1}{2} G^{im} \left[ \frac{\partial G_{km}} {\partial x^j} + \frac{\partial G_{jm}} {\partial x^k} - \frac{\partial G_{jk}} {\partial x^m} \right].
\end{align}
the divergence of a tensor has the form in the curvilinear coordinates
\begin{align} % requires amsmath; align* for no eq. number
 \textrm{div } \mathbf{T}=\left( \frac{1}{\sqrt{G}} \frac{\partial}{\partial x^j} \left(  \sqrt{G}T^{ij}\right) + \Gamma_{mk}^iT^{mk} \right) \mathbf{a}_i.
\end{align}
Noted that
\begin{align} % requires amsmath; align* for no eq. number
   \Gamma_{mk}^iT^{mk} =\sum_{m=1}^3\sum_{k=1}^3\Gamma_{mk}^iT^{mk}
\end{align}

In the appendix \ref{label:appendixA}, we have given the representation in the curvilinear coordinates so that the knowledges of the curvilinear coordinates are defaulted to be known in the following.  In the appendix \ref{label:appendixB}, the geometric formulations such as metric tensor, Christoffel symbol etc. in the spherical polar and cubed-sphere coordinates are presented. Here something is to be cleared.
\begin{itemize}
 \item Coriolis force 
 
 The Coriolis force is defined by
 \begin{align} % requires amsmath; align* for no eq. number
    F_C=-2\mathbf{\Omega} \times \rho \mathbf{u}
 \end{align}
 where $\mathbf{\Omega}$ is the Earth angular velocity pointing from the Earth's origin to pole. It used to adopt the contravariant component to express the vector, thus
 \begin{align} % requires amsmath; align* for no eq. number
    \mathbf{\Omega}&=\omega^1\mathbf{a}_1+\omega^2\mathbf{a}_2+\omega^3\mathbf{a}_3  \nonumber  \\
                               &=\sum_{i=1}^3\omega^i\mathbf{a}_i  = \omega^i\mathbf{a}_i    \quad  \textrm{(Summation used)} \\
                               &=\Omega \cos \varphi \mathbf{e}_\varphi+ \Omega \sin \varphi \mathbf{e}_r   \quad  \textrm{( in the spherical coordinate)}.  \nonumber
 \end{align}
 In fact, due to the same covariant base vector in the radical direction of sphere (as in Eq. \eqref{eq:co_basis3} and \eqref{eq:cubed_co_basis3}), it reads
 \begin{align} % requires amsmath; align* for no eq. number
    \omega^3=\Omega \sin \varphi
 \end{align}
 
  The Coriolis force now has the form
  \begin{align} % requires amsmath; align* for no eq. number
     -2\mathbf{\Omega} \times \rho \mathbf{u} &= -2\rho\omega^j u^k (\mathbf{a}_j \times \mathbf{a}_k)  \\
     &=-2\rho \omega^j u^k e_{ijk} \mathbf{a}^i  \quad \textrm{using Eq. \eqref{eq:identity1}} \\
     &=-2\rho{\sqrt{G}} \left | \begin{array}{ccc}
   \mathbf{a}^1 & \mathbf{a}^2 & \mathbf{a}^3  \\
   \omega^1 & \omega^2 & \omega^3 \\
   u^1 & u^2 & u^3
   \end{array} \right|  \\
   &=-2\rho\sqrt{G} \left[ (\omega^2u^3-\omega^3u^2)\mathbf{a}^1-(\omega^1u^3-\omega^3u^1)\mathbf{a}^2 \right.  \\
   & \quad \quad \left. +(\omega^1u^2-\omega^2u^1)\mathbf{a}^3 \right] \nonumber \quad \textrm{using Eq. \eqref{eq:basevectorrelation_2}}  \nonumber \\
   &=-2\rho\sqrt{G} \left[ (\omega^2u^3-\omega^3u^2)G^{1i}\mathbf{a}_i-(\omega^1u^3-\omega^3u^1)G^{2i}\mathbf{a}_i \right.  \\
   & \quad \quad \left. +(\omega^1u^2-\omega^2u^1)G^{3i}\mathbf{a}_i \right] \nonumber   \\
   &=-2\rho\sqrt{G} \left[ (\omega^2u^3-\omega^3u^2)G^{1i}-(\omega^1u^3-\omega^3u^1)G^{2i} \right.  \\
   & \quad \quad \left. +(\omega^1u^2-\omega^2u^1)G^{3i} \right] \mathbf{a}_i   \quad \textrm{(Summation used)}  \nonumber  \\
   &=F_C^i\mathbf{a}_i
  \end{align}
  where 
  \begin{align} % requires amsmath; align* for no eq. number
     F_C^i = -2\rho\sqrt{G} \left[ (\omega^2u^3-\omega^3u^2)G^{1i}-(\omega^1u^3-\omega^3u^1)G^{2i} +(\omega^1u^2-\omega^2u^1)G^{3i} \right]
  \end{align}

 When the shallow-atmosphere approximations are made, $\mathbf{\Omega} \approx \omega^3 \mathbf{a}_3=(f/2) \mathbf{a}_3$ (where $f=2\Omega \sin \varphi$ called Coriolis parameter) due to only maintaining the vertical component projection of angular velocity $\mathbf{\Omega}$ where it is necessary condition to conserve the energy ($f_\varphi=2\Omega \cos \varphi$ is dropped) \citep{White2005}. In this case  the Coriolis force becomes
\begin{align}
 F_C^i = \rho f\sqrt{G} (-u^1G^{2i}+u^2G^{1i})
\end{align}
 due to $\omega^1=\omega^2=0$.
 
 \item Gravity force
 
 The gravitational source term of the momentum equation has the generic form
 \begin{align} % requires amsmath; align* for no eq. number
    F_G=-\rho g \left(\frac{R}{r}\right)^2 \mathbf{e}_r
 \end{align}
 with radial base vector $\mathbf{e}_r$ in the spherical coordinates and $R$ is the Earth radius. Noted that $\mathbf{a}_3=\mathbf{e}_r$ for radial base vector in the curvilinear coordinates. In the shallow-atmosphere approximations, the gravitational source term yields
 \begin{align} % requires amsmath; align* for no eq. number
    F_G=-\rho g  \mathbf{e}_r
 \end{align} 
 
\end{itemize}

\subsection{Splitting of reference state}
As commonly applied in atmospheric models, the thermodynamic variables are split into a reference state and deviations. The reference state satisfies the stratification balance, i.e the hydrostatic relation in the vertical direction ($z$).  The thermodynamic variables are then written as
\begin{align}\label{eq:linear}
   \rho(\mathbf{x},t)&=\bar{\rho}(\mathbf{x})+\rho'(\mathbf{x},t) \\
    p(\mathbf{x},t)&=\bar{p}(\mathbf{x})+p'(\mathbf{x},t)  \\
   (\rho\theta)(\mathbf{x},t)&=\overline{(\rho\theta)}(\mathbf{x})+(\rho\theta)'(\mathbf{x},t)
\end{align}
where the reference pressure $\bar{p}(r)$ and density $\bar{\rho}(r)$ are in
local hydrostatic balance,
\begin{align}\label{eq:hydrastatic_balance}
   \frac{\partial \bar{p}}{\partial r}=-\bar{\rho}g.
\end{align}

The nonhydrostatic governing equations (Eq. \eqref{eq:curvicoord1}-\eqref{eq:curvicoord4}) with perturbation variables have the form
\begin{align} % requires amsmath; align* for no eq. number
   \frac{\partial \rho'}{\partial t} + \frac{1}{\sqrt{G}} \left[ \frac{\partial (\sqrt{G}\rho u^j)}{\partial x^j} \right] &=0, \label{eq:perturbcurvicoord1} \\
   \frac{\partial \rho u^i}{\partial t} +   \frac{1}{\sqrt{G}}\frac{\partial }{\partial x^j} \left[\sqrt{G}(\rho u^i u^j +G^{ij}p') \right]  &=F^i_H+ F_M^i+F_C^i-\rho' g G^{3i}, \label{eq:perturbcurvicoord2} \\
   \frac{\partial  (\rho \theta)'}{\partial t} + \frac{1}{\sqrt{G}} \left[ \frac{\partial (\sqrt{G}\rho\theta u^j)}{\partial x^j} \right] &=0,  \label{eq:perturbcurvicoord3} \\
     \frac{\partial \rho q_k}{\partial t} + \frac{1}{\sqrt{G}} \left[ \frac{\partial (\sqrt{G}\rho q_k u^j)}{\partial x^j} \right] &=0, \label{eq:perturbcurvicoord4}
\end{align}
where 
\begin{align} % requires amsmath; align* for no eq. number
   F^i_H&=-G^{ij}\frac{\partial \overline{p}}{\partial x^j}  \label{eq:hydrobackgpressuresource} \\
   F_{M}^i&=-\Gamma_{jk}^i(\rho u^ju^k+G^{jk}p') 
\end{align}
is the horizontal variation of the hydrostatic background pressure and the source term due to curvilinear geometry by perturbation pressure, respectively.  In Eq. \eqref{eq:hydrobackgpressuresource}, the nonconservative form of gradient of the hydrostatic background pressure is used while with the conservative form of gradient used for the perturbation pressure, implying from Eq. \eqref{eq:vectorformEq} that 
\begin{align} 
\textrm{grad } p&=\textrm{grad } (\overline{p}+p')  \nonumber \\
         &=\textrm{grad } \overline{p}+\textrm{grad }p' \nonumber \\
         &=\left\{ \left[ G^{ij} \frac{\partial \overline{p}}{\partial x^j} \right]_{\textrm{nonconservative}}+ \left[\frac{1}{\sqrt{G}}\frac{\partial (\sqrt{G}G^{ij}p')}{\partial x^j} + G^{jk}\Gamma_{jk}^i p' \right]_{\textrm{conservative}}\right\}\mathbf{a}_i.
\end{align}

We rewrite the momentum equation in the component form as
\begin{align} % requires amsmath; align* for no eq. number
   \frac{\partial \rho u^1}{\partial t} +   \frac{1}{\sqrt{G}}\frac{\partial }{\partial x^j} \left[\sqrt{G}(\rho u^1 u^j +G^{1j}p') \right]  &=F^1_H+ F_M^1+F_C^1  \\
   \frac{\partial \rho u^2}{\partial t} +   \frac{1}{\sqrt{G}}\frac{\partial }{\partial x^j} \left[\sqrt{G}(\rho u^2 u^j +G^{2j}p') \right]  &=F^2_H+ F_M^2+F_C^2  \\
   \frac{\partial \rho u^3}{\partial t} +   \frac{1}{\sqrt{G}}\frac{\partial }{\partial x^j} \left[\sqrt{G}(\rho u^3 u^j +G^{3j}p') \right]  &=F^3_H+ F_M^3+F_C^3  -\rho' g\left(\frac{R}{r}\right)^2 G^{33}
\end{align}

Note that the linearization of EOS is adopted so that $p'=\epsilon_0(\rho\theta)'$ where $\epsilon_0=R_d^{\gamma}p_0^{-R_d/c_v}\gamma(\overline{\rho\theta})^{\gamma-1}$.

\subsection{The governing equations in the shallow-atmosphere approximation}
In the shallow-atmosphere approximation, $r$ becomes constant and the horizontal projection of the Earth angular velocity is neglected due to conservation of energy \citep{White2005, paul2012}. Let $r=R+z$ ($R$ is the Earth radius) and $dx^3=dr=dz$ where $z$ is the geometry altitude, so the coordinate axes are $(x^1,x^2,x^3)=(\xi,\eta,z)$. Also all $r$ in the geometric tensors in the appendix \ref{label:appendixA} and \ref{label:appendixB} are replaced by the constant $R$. The nonhydrostatic equation of shallow-atmosphere approximation can be recast into
\begin{align} % requires amsmath; align* for no eq. number
   \frac{\partial \rho'}{\partial t} + \frac{1}{\sqrt{G}} \left[ \frac{\partial (\sqrt{G}\rho u^j)}{\partial x^j} \right] &=0, \label{eq:Shallowperturbcurvicoord1} \\
   \frac{\partial \rho u^1}{\partial t} +   \frac{1}{\sqrt{G}}\frac{\partial }{\partial x^j} \left[\sqrt{G}(\rho u^1 u^j +G^{1j}p') \right]  &=F^1_H+ F_M^1+\rho f\sqrt{G} (-u^1G^{21}+u^2G^{11})  \\
   \frac{\partial \rho u^2}{\partial t} +   \frac{1}{\sqrt{G}}\frac{\partial }{\partial x^j} \left[\sqrt{G}(\rho u^2 u^j +G^{2j}p') \right]  &=F^2_H+ F_M^2+\rho f\sqrt{G} (-u^1G^{22}+u^2G^{12})  \\
 \frac{\partial \rho u^3}{\partial t} +   \frac{1}{\sqrt{G}}\frac{\partial }{\partial x^j} \left[\sqrt{G}(\rho u^3 u^j +G^{3j}p') \right] 
 &= -\rho' g \\
   \frac{\partial  (\rho \theta)'}{\partial t} + \frac{1}{\sqrt{G}} \left[ \frac{\partial (\sqrt{G}\rho\theta u^j)}{\partial x^j} \right] &=0,  \label{eq:Shallowperturbcurvicoord5} \\
     \frac{\partial \rho q_k}{\partial t} + \frac{1}{\sqrt{G}} \left[ \frac{\partial (\sqrt{G}\rho q_k u^j)}{\partial x^j} \right] &=0, \label{eq:Shallowperturbcurvicoord6}
\end{align}
where 
\begin{align} % requires amsmath; align* for no eq. number
    F_M^{i(=1,2)} &=\left( 
    \begin{array}{c}
    -\Gamma_{jk}^1\rho u^ju^k \\
    -\Gamma_{jk}^2\rho u^ju^k\\
    \end{array}
    \right)    \quad\quad \textrm{ for the cubed-sphere coordinates}   \\
 &=\left( 
    \begin{array}{c}
    -\Gamma_{jk}^1\rho u^ju^k \\
    -\Gamma_{jk}^2\rho u^ju^k-p'\tan\varphi/R^2 \\
    \end{array}
    \right)  \quad   \textrm{ for the spherical polar coordinates}   
\end{align}
and 
\begin{align} % requires amsmath; align* for no eq. number
   F^{i(=1,2)}_H=\left( 
    \begin{array}{c}
    -\left(G^{11}\frac{\partial \overline{p}}{\partial x^1}+G^{12}\frac{\partial \overline{p}}{\partial x^2}\right)  \\
    -\left(G^{21}\frac{\partial \overline{p}}{\partial x^1}+G^{22}\frac{\partial \overline{p}}{\partial x^2}\right) \\
    \end{array}
    \right).
\end{align}

Now we obtain the specific equations of the shallow-atmosphere approximation in the general curvilinear coordinates for the spherical polar and cubed-sphere systems:
\begin{enumerate}
\item Mass equation 
\begin{align} % requires amsmath; align* for no eq. number
   \frac{\partial \rho'}{\partial t} + \frac{1}{\sqrt{G}} \left[ \frac{\partial (\sqrt{G}\rho u^\xi)}{\partial \xi} +\frac{\partial (\sqrt{G}\rho u^\eta)}{\partial \eta} +\frac{\partial (\sqrt{G}\rho w)}{\partial z} \right] &=0
\end{align}

\item $u^{\xi}$-momentum equation
\begin{align} % requires amsmath; align* for no eq. number
     & \frac{\partial \rho u^\xi}{\partial t} +   \frac{1}{\sqrt{G}} \left\{  \frac{\partial }{\partial \xi} \left[\sqrt{G}(\rho u^\xi u^\xi +G^{11}p') \right] + \frac{\partial }{\partial \eta} \left[\sqrt{G}(\rho u^\xi u^\eta +G^{12}p') \right]+  \frac{\partial }{\partial z} \left[\sqrt{G}(\rho u^\xi w ) \right]\right\}  \nonumber  \\
    &= -\left(G^{11}\frac{\partial \overline{p}}{\partial \xi}+G^{12}\frac{\partial \overline{p}}{\partial \eta}\right) - \left( \Gamma_{11}^1\rho u^\xi u^\xi+ 2\Gamma_{12}^1\rho u^\xi u^\eta +\Gamma_{22}^1\rho u^\eta u^\eta \right) +\rho f\sqrt{G} (-u^\xi G^{21}+u^\eta G^{11})
\end{align}

\item $u^\eta$-momentum equation
\begin{align} % requires amsmath; align* for no eq. number
     & \frac{\partial \rho u^\eta}{\partial t} +   \frac{1}{\sqrt{G}} \left\{  \frac{\partial }{\partial \xi} \left[\sqrt{G}(\rho u^\eta u^\xi +G^{21}p') \right] + \frac{\partial }{\partial \eta} \left[\sqrt{G}(\rho u^\eta u^\eta +G^{22}p') \right]+  \frac{\partial }{\partial z} \left[\sqrt{G}(\rho u^\eta w ) \right]\right\}  \nonumber  \\
    &= -\left(G^{21}\frac{\partial \overline{p}}{\partial \xi}+G^{22}\frac{\partial \overline{p}}{\partial \eta}\right) - \left( \Gamma_{11}^2 \rho u^\xi u^\xi+ 2\Gamma_{12}^2\rho u^\xi u^\eta +\Gamma_{22}^2\rho u^\eta u^\eta +\delta_p \right) \nonumber \\
    &+\rho f\sqrt{G} (-u^\xi G^{22}+u^\eta G^{12})
\end{align}

\item $w$-momentum equation
\begin{align} % requires amsmath; align* for no eq. number
      \frac{\partial \rho w}{\partial t} + \frac{1}{\sqrt{G}}\left\{ \frac{\partial }{\partial \xi} (\sqrt{G}\rho w u^\xi) + \frac{\partial }{\partial \eta} (\sqrt{G}\rho w u^\eta)+ \frac{\partial }{\partial z} \left[\sqrt{G}(\rho ww +p') \right] \right\}= -\rho' g
\end{align}

\item Potential temperature equation
\begin{align} % requires amsmath; align* for no eq. number
   \frac{\partial (\rho\theta)'}{\partial t} + \frac{1}{\sqrt{G}} \left[ \frac{\partial (\sqrt{G}\rho\theta u^\xi)}{\partial \xi} +\frac{\partial (\sqrt{G}\rho\theta u^\eta)}{\partial \eta} +\frac{\partial (\sqrt{G}\rho\theta w)}{\partial z} \right] &=0
\end{align}

\item The $k$th-moisture$/$tracer equation
\begin{align} % requires amsmath; align* for no eq. number
   \frac{\partial (\rho q_k)}{\partial t} + \frac{1}{\sqrt{G}} \left[ \frac{\partial (\sqrt{G}\rho q_k u^\xi)}{\partial \xi} +\frac{\partial (\sqrt{G}\rho q_k u^\eta)}{\partial \eta} +\frac{\partial (\sqrt{G}\rho q_k w)}{\partial z} \right] &=0
\end{align}

In the $u^\eta$-momentum equation $\delta_p$ has the value
\begin{align} % requires amsmath; align* for no eq. number
   \delta_p = \left\{
   \begin{array}{ll}
     0,  &\textrm{for the cubed-sphere coordinates}  \\
     p'\tan\varphi/R^2, & \textrm{for the spherical polar coordinates}  \\
   \end{array}
   \right.
\end{align}
\end{enumerate}
Here we have replaced the superscript indices $(1,2,3)$ by $(\xi,\eta,z)$ when representing the contravariant velocity and the coordinate axes $(x^1,x^2,x^3)$ by $(\xi,\eta,z)$ . Obviously, the horizontal curvilinear coordinates become the spherical polar system when $(\xi,\eta)=(\lambda,\varphi)$ while they are the cubed-sphere coordinates if $(\xi,\eta)=(\alpha,\beta)$.

\subsection{Governing equations with the effects of topography}
In the presence of topography, the height-based terrain-following coordinate introduced by \cite{Galchen1975}
is utilized to map the physical space $(x^1,x^2,z)=(\xi,\eta,z)$ into the computational domain $(x^1,x^2,\zeta)=(\xi,\eta,\zeta)$ via the transformation relationship $\zeta=\zeta(x^1,x^2,z)=\zeta(\xi,\eta,z)$. Before the transformation of vertical coordinate, we should have the vision of separating the 3D curvilinear coordinates into 2D spherical/cubed-sphere surface coordinate ($(\mathbf{a}_1,\mathbf{a}_2)$) and radial ccordinate ($\mathbf{a}_3$). For convenience, we denote the curvilinear geometric quantities as $(\cdot)_c$ and the geometric quantities of vertical coordinate transformation as $(\cdot)_v$. The total Jacobian of transformation can be derived in the following manner: the first metric Jacobian associated with spherical/cubed-sphere transformation $(\xi,\eta,z)$ is $\sqrt{G}_c$; the second Jacobian with the vertical transformation $z \rightarrow \zeta$ has $\sqrt{G}_v$. Therefore, the final composite Jacobian of transformation is $\sqrt{G}=\sqrt{G}_c\sqrt{G}_v$. The conservation form by the chain rule \citep{Clark1977} can be expressed in the vertical coordinate transform as 
\begin{align} % requires amsmath; align* for no eq. number
  \sqrt{G}_v\left( \frac{\partial \phi} {\partial \xi}\right)_z &=  \frac{\partial } {\partial \xi}\left(\sqrt{G}_v \phi\right)_\zeta + \frac{\partial (\sqrt{G}_vG^{13}_v\phi)}{\partial \zeta} \\
   \sqrt{G}_v\left( \frac{\partial \phi} {\partial \eta}\right)_z &= \frac{\partial } {\partial \eta}\left(\sqrt{G}_v \phi\right)_\zeta + \frac{\partial (\sqrt{G}_vG^{23}_v\phi)}{\partial \zeta}  \\
   \sqrt{G}_v\left( \frac{\partial \phi} {\partial z}\right)_z &=  \frac{\partial \phi}{\partial \zeta} 
\end{align}
where 
\begin{align} % requires amsmath; align* for no eq. number
   \sqrt{G}_v=\frac{\partial z}{\partial \zeta}, G_v^{13}=\frac{\partial \zeta}{\partial \xi}, G_v^{23}=\frac{\partial \zeta}{\partial \eta}
\end{align}

 From the definition of the transformed vertical velocity, we have
\begin{align} % requires amsmath; align* for no eq. number
   \tilde{w} =\frac{d \zeta}{d t} &=\frac{\partial \zeta}{\partial t}+ u^1\frac{\partial \zeta}{\partial x^1}+u^2\frac{\partial \zeta}{\partial x^2}+u^3\frac{\partial \zeta}{\partial x^3}  \\
                  &=u^\xi\frac{\partial \zeta}{\partial \xi}+u^\eta\frac{\partial \zeta}{\partial \eta}+w\frac{\partial \zeta}{\partial z} \label{eq:w_trans}   \\
                  &=G^{13}_vu^\xi+G^{23}_vu^\eta+\frac{1}{\sqrt{G}_v}w  \\
                  &=\frac{1}{\sqrt{G}_v}\left(w+ \sqrt{G}_vG^{13}_vu^\xi+\sqrt{G}_vG^{23}_vu^\eta \right)  \label{eq:w-eq}
\end{align}

Now we obtain the transformed governing equation as follows
\begin{itemize}
\item Mass equation
\begin{align} % requires amsmath; align* for no eq. number
   \frac{\partial \rho'}{\partial t} + \frac{1}{\sqrt{G}} \left[ \frac{\partial (\sqrt{G}\rho u^\xi)}{\partial \xi} +\frac{\partial (\sqrt{G}\rho u^\eta)}{\partial \eta} +\frac{\partial (\sqrt{G}\rho \tilde{w})}{\partial \zeta} \right] &=0
\end{align}

\item $u^{\xi}$-momentum equation
\begin{align} % requires amsmath; align* for no eq. number
     & \frac{\partial \rho u^\xi}{\partial t} +   \frac{1}{\sqrt{G}} \left\{  \frac{\partial }{\partial \xi} \left[\sqrt{G}(\rho u^\xi u^\xi +G^{11}_c p') \right] + \frac{\partial }{\partial \eta} \left[\sqrt{G}(\rho u^\xi u^\eta +G^{12}_cp') \right]+ \right. \nonumber \\
  &\left.    \frac{\partial }{\partial \zeta} \left[\sqrt{G}(\rho u^\xi \tilde{w} + {G}_v^{13} {G}_c^{11}p'+{G}_v^{23} {G}_c^{12}p') \right]\right\}  \nonumber  \\
    &= F_H^1 + F_M^1 +\rho f\sqrt{G}_c (-u^\xi G^{21}_c+u^\eta G^{11}_c)
\end{align}

\item $u^\eta$-momentum equation
\begin{align} % requires amsmath; align* for no eq. number
     & \frac{\partial \rho u^\eta}{\partial t} +   \frac{1}{\sqrt{G}} \left\{  \frac{\partial }{\partial \xi} \left[\sqrt{G}(\rho u^\eta u^\xi +G^{21}_c p') \right] + \frac{\partial }{\partial \eta} \left[\sqrt{G}(\rho u^\eta u^\eta +G^{22}_c p') \right] + \right.  \nonumber \\
   &  \left.  \frac{\partial }{\partial \zeta} \left[\sqrt{G}(\rho u^\eta \tilde{w} + {G}_v^{13} {G}_c^{21}p'+{G}_v^{23} {G}_c^{22}p' ) \right]\right\}  \nonumber  \\
    &= F_H^2+F_M^2+\rho f\sqrt{G}_c (-u^\xi G^{22}_c+u^\eta G^{12}_c)
\end{align}

\item $w$-momentum equation
\begin{align} % requires amsmath; align* for no eq. number
      \frac{\partial \rho w}{\partial t} + \frac{1}{\sqrt{G}}\left\{ \frac{\partial }{\partial \xi} (\sqrt{G}\rho w u^\xi) + \frac{\partial }{\partial \eta} (\sqrt{G}\rho w u^\eta)+ \frac{\partial }{\partial \zeta} \left[\sqrt{G}\rho w\tilde{w} +\sqrt{G}_c p') \right] \right\}= -\rho' g
  % (\sqrt{G}\rho w u^\eta)+ \frac{\partial }{\partial \zeta} \left[\sqrt{G}(\rho w\tilde{w} +p') \right] \right\}= -\rho' g
\end{align}

\item Potential temperature equation
\begin{align} % requires amsmath; align* for no eq. number
   \frac{\partial (\rho\theta)'}{\partial t} + \frac{1}{\sqrt{G}} \left[ \frac{\partial (\sqrt{G}\rho\theta u^\xi)}{\partial \xi} +\frac{\partial (\sqrt{G}\rho\theta u^\eta)}{\partial \eta} +\frac{\partial (\sqrt{G}\rho\theta \tilde{w})}{\partial \zeta} \right] &=0
\end{align}

\item The $k$th-moisture$/$tracer equation
\begin{align} % requires amsmath; align* for no eq. number
   \frac{\partial (\rho q_k)}{\partial t} + \frac{1}{\sqrt{G}} \left[ \frac{\partial (\sqrt{G}\rho q_k u^\xi)}{\partial \xi} +\frac{\partial (\sqrt{G}\rho q_k u^\eta)}{\partial \eta} +\frac{\partial (\sqrt{G}\rho q_k \tilde{w})}{\partial \zeta} \right] &=0
\end{align}
where 
\begin{align} % requires amsmath; align* for no eq. number
   F_H^1&=-\left \{ \frac{G^{11}_c}{\sqrt{G}_v} \left( \frac{\partial (\sqrt{G}_v \overline{p})}{\partial \xi}+\frac{\partial (\sqrt{G}_vG_v^{13}\overline{p})}{\partial \zeta}\right)+\frac{G^{12}_c}{\sqrt{G}_v}\left(\frac{\partial (\sqrt{G}_v \overline{p})}{\partial \eta} + \frac{\partial (\sqrt{G}_vG_v^{23}\overline{p})}{\partial \zeta} \right)\right \}  \\
   F_H^2&=-\left\{ \frac{G^{21}_c}{\sqrt{G}_v} \left(\frac{\partial (\sqrt{G}_v\overline{p})}{\partial \xi}+\frac{\partial (\sqrt{G}_vG_v^{13}\overline{p})}{\partial \zeta} \right)+\frac{G^{22}_c}{\sqrt{G}_v} \left( \frac{\partial (\sqrt{G}_v\overline{p})}{\partial \eta}+ \frac{\partial (\sqrt{G}_vG_v^{23}\overline{p})}{\partial \zeta} \right)\right\}
\end{align}
\begin{align} % requires amsmath; align* for no eq. number
   F_M^1&=-\left( \Gamma_{11}^1\rho u^\xi u^\xi+ 2\Gamma_{12}^1\rho u^\xi u^\eta +\Gamma_{22}^1\rho u^\eta u^\eta \right)  \\
   F_M^2&=- \left( \Gamma_{11}^2 \rho u^\xi u^\xi+ 2\Gamma_{12}^2\rho u^\xi u^\eta +\Gamma_{22}^2\rho u^\eta u^\eta +\delta_p \right)
\end{align}

\end{itemize}

The above nonhydrostatic governing equations can be written in the compact flux form
\begin{align} % requires amsmath; align* for no eq. number
   \frac{\partial \mathbf{q}}{\partial t} +\frac{\partial \mathbf{f}(\mathbf{q})}{\partial \xi} +\frac{\partial \mathbf{g}(\mathbf{q})}{\partial \eta} + \frac{\partial \mathbf{h}(\mathbf{q})}{\partial \zeta} =\mathbf{S}(\mathbf{q})
\end{align}
where
\begin{align} % requires amsmath; align* for no eq. number
   \mathbf{q}=(\sqrt{G}\rho',\sqrt{G}\rho u^\xi, \sqrt{G}\rho u^\eta, \sqrt{G}\rho w,\sqrt{G}(\rho\theta)',\sqrt{G}\rho q_k)^T
\end{align}
\begin{align} % requires amsmath; align* for no eq. number
   \mathbf{f}(\mathbf{q})=\left( 
   \begin{array}{c}
       \sqrt{G}\rho u^\xi  \\
       \sqrt{G}(\rho u^\xi u^\xi +G^{11}_c p')  \\
       \sqrt{G}(\rho u^\eta u^\xi +G^{21}_c p') \\
       \sqrt{G}\rho w u^\xi \\
       \sqrt{G}\rho\theta u^\xi \\
       \sqrt{G}\rho q_k u^\xi
   \end{array}
   \right)
\end{align}
\begin{align} % requires amsmath; align* for no eq. number
   \mathbf{g}(\mathbf{q})=\left( 
   \begin{array}{c}
       \sqrt{G}\rho u^\eta  \\
       \sqrt{G}(\rho u^\xi u^\eta +G^{12}_cp') \\
       \sqrt{G}(\rho u^\eta u^\eta +G^{22}_c p')  \\
       \sqrt{G}\rho w u^\eta  \\
       \sqrt{G}\rho\theta u^\eta \\
       \sqrt{G}\rho q_k u^\eta
   \end{array}
   \right)
\end{align}
\begin{align} % requires amsmath; align* for no eq. number
   \mathbf{h}(\mathbf{q})=\left( 
   \begin{array}{c}
       \sqrt{G}\rho \tilde{w}  \\
       \sqrt{G}(\rho u^\xi \tilde{w} + {G}_v^{13} {G}_c^{11}p'+{G}_v^{23} {G}_c^{12}p') \\
       \sqrt{G}(\rho u^\eta \tilde{w} + {G}_v^{13} {G}_c^{21}p'+{G}_v^{23} {G}_c^{22}p' ) \\
       \sqrt{G}\rho w\tilde{w} +\sqrt{G}_cp' \\
%       \sqrt{G}(\rho w\tilde{w} +p') \\       
       \sqrt{G}\rho\theta \tilde{w} \\
       \sqrt{G}\rho q_k \tilde{w} 
   \end{array}
   \right)
\end{align}
\begin{align} % requires amsmath; align* for no eq. number
   \mathbf{S}(\mathbf{q})=\sqrt{G}\left( 
   \begin{array}{c}
        0 \\
        F_H^1 + F_M^1 +\rho f\sqrt{G}_c (-u^\xi G^{21}_c+u^\eta G^{11}_c) \\
        F_H^2+F_M^2+\rho f\sqrt{G}_c (-u^\xi G^{22}_c+u^\eta G^{12}_c) \\
        -\rho' g \\
        0 \\
        0
   \end{array}
   \right)
\end{align}

\subsection{Flux Jacobian }
The flux Jacobian in the $x$-direction is given by
\begin{align} 
\mathbf{A}=\frac{\partial \mathbf{f}}{\partial \mathbf{q}}=\left( 
   \begin{array}{ccccc}
    0 & 1 & 0 & 0 & 0 \\
    -u^{\xi} u^{\xi} & 2u^{\xi}  & 0 & 0 & G_{c}^{11} \epsilon_0  \\
    -u^{\xi} u^{\eta} & u^{\eta}  & u^{\xi} & 0 & G_{c}^{21} \epsilon_0  \\    
    -u^{\xi} w & w  & 0 & u^{\xi} & 0  \\   
    -u^{\xi} \theta & \theta  & 0 & 0 & u^{\xi}   \\       
   \end{array}
   \right).
\end{align}
The eigenvalues are $(u^{\xi}-\sqrt{G_{c}^{11}}a,u^{\xi},u^{\xi},u^{\xi},u^{\xi}+\sqrt{G_{c}^{11}}a)$ where $a=\sqrt{\epsilon_0\theta}$ is sound speed, and the corresponding right eigenvectors $\mathbf{R}_x$ and left eigenvectors $\mathbf{L}_x$, respectively, are
\begin{align} 
\mathbf{R}_x=\left( 
   \begin{array}{ccccc}
    1 & 1 & 0 & 0 & 1 \\
    u^{\xi} - \sqrt{G_{c}^{11}}a & u^{\xi}  & 0 & 0 & u^{\xi} + \sqrt{G_{c}^{11}}a  \\
    \frac{\sqrt{G_{c}^{11}}u^{\eta}-G_{c}^{21}a}{\sqrt{G_{c}^{11}}} & 0  & 1 & 0 & \frac{\sqrt{G_{c}^{11}}u^{\eta}+G_{c}^{21}a}{\sqrt{G_{c}^{11}}} \\    
    w & 0  & 0 & 1 & w  \\   
     \theta & 0  & 0 & 0 & \theta   \\       
   \end{array}
   \right)
\end{align}
and 
\begin{align} 
\mathbf{L}_x=\mathbf{R}^{-1}_x=\left( 
   \begin{array}{ccccc}
    \frac{u^{\xi}}{2\sqrt{G_{c}^{11}}a} & -\frac{1}{2\sqrt{G_{c}^{11}}a} & 0 & 0 & \frac{1}{2\theta} \\
    1 & 0  & 0 & 0 & -\frac{1}{\theta}  \\
    \frac{G_{c}^{21}u^{\xi}}{G_{c}^{11}} & -\frac{G_{c}^{21}}{G_{c}^{11}}  & 1 & 0 & -\frac{u^{\eta}}{\theta}\\    
    0 & 0  & 0 & 1 & -\frac{w}{\theta}  \\   
    -\frac{u^{\xi}}{2\sqrt{G_{c}^{11}}a} & \frac{1}{2\sqrt{G_{c}^{11}}a} & 0 & 0 & \frac{1}{2\theta}     
   \end{array}
   \right).
\end{align}

The flux Jacobian in the $y$-direction reads
\begin{align} 
\mathbf{B}=\frac{\partial \mathbf{g}}{\partial \mathbf{q}}=\left( 
   \begin{array}{ccccc}
    0 & 0 & 1 & 0 & 0 \\
    -u^{\xi} u^{\eta} & u^{\eta}  & u^{\xi} & 0 & G_{c}^{12} \epsilon_0  \\
    -u^{\eta} u^{\eta} & 0  & 2u^{\eta} & 0 & G_{c}^{22} \epsilon_0  \\    
    -u^{\eta} w & 0  & w & u^{\eta} & 0  \\   
    -u^{\eta} \theta & 0  & \theta & 0 & u^{\eta}   \\       
   \end{array}
   \right)
\end{align}
The eigenvalues are $(u^{\eta}-\sqrt{G_{c}^{22}}a,u^{\eta},u^{\eta},u^{\eta},u^{\eta}+\sqrt{G_{c}^{22}}a)$ where $a=\sqrt{\epsilon_0\theta}$ is sound speed, and the corresponding right eigenvectors $\mathbf{R}_y$ and left eigenvectors $\mathbf{L}_y$, respectively, are
\begin{align} 
\mathbf{R}_y=\left( 
   \begin{array}{ccccc}
    1 & 0 & 1 & 0 & 1 \\
        \frac{\sqrt{G_{c}^{22}}u^{\xi}-G_{c}^{21}a}{\sqrt{G_{c}^{22}}} & 0  & 0 & 1 & \frac{\sqrt{G_{c}^{22}}u^{\xi}+G_{c}^{21}a}{\sqrt{G_{c}^{22}}} \\  
    u^{\eta} - \sqrt{G_{c}^{22}}a  & 0 & u^{\eta} & 0 & u^{\eta} + \sqrt{G_{c}^{22}}a  \\
    w & 1  & 0 & 0 & w  \\   
     \theta & 0  & 0 & 0 & \theta   \\       
   \end{array}
   \right)
\end{align}
and
\begin{align} 
\mathbf{L}_y=\mathbf{R}^{-1}_y=\left( 
   \begin{array}{ccccc}
    \frac{u^{\eta}}{2\sqrt{G_{c}^{22}}a} & 0 & -\frac{1}{2\sqrt{G_{c}^{22}}a}  & 0 & \frac{1}{2\theta} \\
    0 & 0  & 0 & 1 & -\frac{w}{\theta}  \\
    1 & 0  & 0 & 0 & -\frac{1}{\theta}  \\   
    \frac{G_{c}^{12}u^{\eta}}{G_{c}^{22}} &  1 & -\frac{G_{c}^{12}}{G_{c}^{22}}  & 0 & -\frac{u^{\xi}}{\theta}\\    
    -\frac{u^{\eta}}{2\sqrt{G_{c}^{22}}a} &  0 & \frac{1}{2\sqrt{G_{c}^{22}}a} & 0 & \frac{1}{2\theta}     
   \end{array}
   \right).
\end{align}

The flux Jacobian in the $\zeta$-direction reads
\begin{align} 
\mathbf{C}=\frac{\partial \mathbf{h}}{\partial \mathbf{q}}=\left( 
   \begin{array}{ccccc}
    0 & \frac{G_{VX}}{\sqrt{G}_v} & \frac{G_{VY}}{\sqrt{G}_v} & \frac{1}{\sqrt{G}_v} & 0 \\
    -u^{\xi} \tilde{w} & \tilde{w}+\frac{G_{VX}u^{\xi} }{\sqrt{G}_v} & \frac{G_{VY} u^{\xi} }{\sqrt{G}_v} &  \frac{u^{\xi}}{\sqrt{G}_v} & G_{X} \epsilon_0  \\
    -u^{\eta} \tilde{w} & \frac{G_{VX}u^{\eta} }{\sqrt{G}_v}  &  \tilde{w}+\frac{G_{VY}u^{\eta} }{\sqrt{G}_v}  & \frac{u^{\eta}}{\sqrt{G}_v} & G_{Y} \epsilon_0  \\    
    -w \tilde{w} & \frac{G_{VX}w }{\sqrt{G}_v}   & \frac{G_{VY}w }{\sqrt{G}_v} & \tilde{w}+\frac{w }{\sqrt{G}_v} & \frac{\epsilon_0}{\sqrt{G}_v}  \\   
    -\theta \tilde{w} & \frac{G_{VX}\theta }{\sqrt{G}_v}  & \frac{G_{VY}\theta }{\sqrt{G}_v} & \frac{\theta }{\sqrt{G}_v}  & \tilde{w}   \\       
   \end{array}
   \right)
\end{align}
where $G_{VX}=\sqrt{G}_vG_v^{13}$, $G_{VY}=\sqrt{G}_vG_v^{23}$, $G_X=G_v^{13}G_c^{11}+G_v^{23}G_c^{12}$ and $G_Y=G_v^{13}G_c^{21}+G_v^{23}G_c^{22}$.  The eigenvalues are $\left(\tilde{w},\tilde{w},\tilde{w},\tilde{w}-\frac{\sqrt{M}}{\sqrt{G}_v}a,\tilde{w}+\frac{\sqrt{M}}{\sqrt{G}_v}a \right)$ where $a=\sqrt{\epsilon_0\theta}$ is sound speed and $M=1+\sqrt{G}_vG_{VX}G_X+\sqrt{G}_vG_{VY}G_Y$. The corresponding right eigenvectors $\mathbf{R}_\zeta$ and left eigenvectors $\mathbf{L}_\zeta$, respectively, are
\begin{align} 
\mathbf{R}_\zeta=\left( 
   \begin{array}{ccccc}
    \frac{1}{\sqrt{G}_v\tilde{w}} & \frac{G_{VY}}{\sqrt{G}_v\tilde{w}} & \frac{G_{VX}}{\sqrt{G}_v\tilde{w}} & 1 & 1 \\
    0 & 0 & 1 & u^{\xi}-G_X a \frac{\sqrt{G}_v}{\sqrt{M}}  & u^{\xi}+G_X a \frac{\sqrt{G}_v}{\sqrt{M}}   \\
    0 & 1 & 0 & u^{\eta}-G_Y a \frac{\sqrt{G}_v}{\sqrt{M}}   & u^{\eta}+G_Y a \frac{\sqrt{G}_v}{\sqrt{M}}   \\    
    1 & 0 & 0 & w-\frac{a}{\sqrt{M}}  & w+\frac{a}{\sqrt{M}}  \\   
     0 & 0  & 0 & \theta & \theta   \\       
   \end{array}
   \right)
\end{align}
and
\begin{align} 
\mathbf{L}_\zeta=\mathbf{R}^{-1}_\zeta=\left( 
   \begin{array}{ccccc}
    \frac{\sqrt{G}_v\tilde{w}}{M} & -\frac{G_{VX}}{M} & -\frac{G_C G_{VY}}{M}  & 1-\frac{1}{M} & -\frac{w}{\theta} \\
    \frac{G_Y\sqrt{G}_v^2\tilde{w}}{M} & -\frac{G_{VX}G_Y\sqrt{G}_v}{M}  & \frac{1+G_{VX}G_X\sqrt{G}_v}{M} & -\frac{G_Y\sqrt{G}_v}{M} & -\frac{u^{\eta}}{\theta}  \\
    \frac{G_X\sqrt{G}_v^2\tilde{w}}{M} &  \frac{1+G_{VY}G_Y\sqrt{G}_v}{M}  & -\frac{G_{VY}G_X\sqrt{G}_v}{M} & -\frac{G_X\sqrt{G}_v}{M} & -\frac{u^{\xi}}{\theta}  \\ 
    \frac{\sqrt{G}_v }{2a \sqrt{M}}\tilde{w} &  -\frac{G_{VX}}{2a\sqrt{M}} & -\frac{G_{VY}}{2a\sqrt{M}}  & -\frac{1}{2a\sqrt{M}} & \frac{1}{2\theta}\\    
    -\frac{\sqrt{G}_v }{2a \sqrt{M}}\tilde{w} & \frac{G_{VX}}{2a\sqrt{M}}  & \frac{G_{VY}}{2a\sqrt{M}} & \frac{1}{2a\sqrt{M}} & \frac{1}{2\theta}     
   \end{array}
   \right).
\end{align}

\section{summary}
Based on differential geometry approach, we have derived the flux-form atmospheric governing equation in the the general curvilinear coordinate system which are currently utilized in a high-order nonhydrostatic MCV dynamical core.  The explicit flux-form atmospheric governing equations  in the shallow-atmosphere approximation are given. In the unified MCV dynamical core framework, the horizontal curvilinear coordinates become the spherical polar system when $(\xi,\eta)=(\lambda,\varphi)$ where $\lambda$ and $\varphi$ represent the longitude and latitude directions, while they are the cubed-sphere coordinates if $(\xi,\eta)=(\alpha,\beta)$ where $\alpha$ and $\beta$ denote the cube coordinates. Of course, the coordinate system directly reduces to Cartesian coordinate when $(\xi,\eta)=(x,y)$ where $x$ and $y$ represent the natural coordinate. It is noted that the projection metric tensors like spherical polar system and Cartesian coordinate become simple due to orthogonal properties of coordinate. In the current nonhydrostatic MCV framework, it is easy to switch to one of the coordinate systems: spherical polar system, cubed-sphere system and Cartesian system by simply changing the projection relations. In addition, the flux Jacobian of the three coordinate system are given in this manuscript.

%%%%%%%%%%%%%%%%%%%%%%%%%%%%%%%%%%%%%%%%%%%%%%%%%%%%%%%%%%%%%%%%%%%%%
% Create a bibliography directory and place your .bib file there.
%%%%%%%%%%%%%%%%%%%%%%%%%%%%%%%%%%%%%%%%%%%%%%%%%%%%%%%%%%%%%%%%%%%%%
{\clearpage}
\bibliographystyle{./ametsoc}
%\bibliography{./references}

%%%%%%%%%%%%%%%%%%%%%%%%%%%%%%%%%%%%%%%%%%%%%%%%%%%%%%%%%%%%%%%%%%%%%
% APPENDIXES
%%%%%%%%%%%%%%%%%%%%%%%%%%%%%%%%%%%%%%%%%%%%%%%%%%%%%%%%%%%%%%%%%%%%%

\clearpage
%%%%%%%%%%%%%%
% appendix setting  by XL LI
%%%%%%%%%%%%%%
%\appendix\newpage\markboth{Appendix}{Appendix}
%\renewcommand{\thesection}{\Alph{section}}
{\renewcommand\theequation{A\arabic{equation}}%
\setcounter{equation}{0}% reset counter

\begin{center}
\bf \Huge Appendix
\end{center}

\appendix
\section{Representation in curvilinear coordinates}  \label{label:appendixA}
In general curvilinear coordinate $x^i(i=1,2,3)$, unlike the Cartesian coordinate, the base vectors are not constants either in magnitude or direction. Here we use the standard practice to express the representation in curvilinear coordinates \citep{Thompson1985,Warsi2005}.

\subsection{Base Vectors}
\begin{itemize}
\item {\em{Covariant base vectors}} are defined by
\begin{align} % requires amsmath; align* for no eq. number
   \mathbf{a}_i=\frac{\partial \mathbf{r}}{\partial x^i}
\end{align}
\item {\em{Contravariant base vectors}} are defined by
\begin{align} % requires amsmath; align* for no eq. number
   \mathbf{a}^i=\nabla x^i
\end{align}
\end{itemize}
where $\mathbf{r}$ is the position vector. They have the relation as follows
\begin{align} % requires amsmath; align* for no eq. number
   \mathbf{a}^i \cdot \mathbf{a}_j=\delta^i_j, \quad i,j,=1,2,3   \label{eq:KroSym}
\end{align}
where $\delta^i_j$ is the \emph{Kronecker symbol}
\begin{align} % requires amsmath; align* for no eq. number
   \delta^i_j&=1  \quad \text{if}  \quad i=j  \nonumber \\
                  &=0 \quad \text{if} \quad i\neq j
\end{align}

\begin{itemize}
\item Scale Factors

The covariant and contravariant base vectors are not unit vectors. In particular, consider the covariant base vectors and introduce the \emph{unit} triad $\mathbf{\hat{a}}_i$, with $|\mathbf{a}_i|=\sqrt{\mathbf{a}_i \cdot \mathbf{a}_i}$
\begin{align} % requires amsmath; align* for no eq. number
   \mathbf{\hat{a}}_i = \frac{\mathbf{a}_i}{|\mathbf{a}_i|}=\frac{\mathbf{a}_i}{\sqrt{G_{ii}}}    \quad \textrm{(no summation)}
\end{align}
The lengths of the covariant base vectors are usually denoted by $h$ and are called the \emph{scale factors}
\begin{align} % requires amsmath; align* for no eq. number
   h_i=|\mathbf{a}_i| = \sqrt{G_{ii}}    \quad \textrm{(no summation)}
\end{align}
\end{itemize}

\subsection{Vector and Tensor}
\begin{itemize}
\item In terms of the two base vectors, a vector $\mathbf{u}$ can now be expressed by using the summation convention as
\begin{align} % requires amsmath; align* for no eq. number
   \mathbf{u}&=u^i \mathbf{a}_i,   \\
                   &=u_i  \mathbf{a}^i,
\end{align}
where $u^i$ and $u_i$ are the so-called \emph{contravariant} and \emph{covariant} component of a vector.
\item  Parallel and orthogonal projections
\begin{align} % requires amsmath; align* for no eq. number
   u_i =\mathbf{u}\cdot\mathbf{a}_i, \\
   u^i =\mathbf{u}\cdot\mathbf{a}^i .
\end{align}
From the viewing of projection, they are the parallel and orthogonal  projections of a vector.
\end{itemize}
Similarly a second-order tensor $\mathbf{T}$ (or $\overleftrightarrow{\mathbf{T}}$) is now represented as
\begin{align} % requires amsmath; align* for no eq. number
   \mathbf{T}&=T^{ij}\mathbf{a}_i\mathbf{a}_j  \\
                    &=T_{ij}\mathbf{a}^i\mathbf{a}^j  \\
                    &=T^i_j\mathbf{a}_i\mathbf{a}^j =T_j^{\text{  } i}\mathbf{a}^j\mathbf{a}_i
\end{align}
where $T^{ij}$ and $T_{ij}$  are the contravariant and covariant components of $\mathbf{T}$, respectively, and the superscript index $i$ stands for the contravariant and subscript $j$ stands for the covariant nature of $T^i_j$ and $T_j^{\text{  } i}$.

\begin{itemize}
\item Physical Components of a Vector

The contravariant and covariant components of a vector do not have the same physical significance in a curvilinear coordinate system as they do in a rectangular Cartesian system; accutally they often have different dimensions. For instance, the increment of a position vector $\mathbf{r}$ has the contravariant components $(dr,d\theta,dz)$ in cylindrical coordinates
\begin{align} % requires amsmath; align* for no eq. number
   d \mathbf{r}=d r \mathbf{a}_1+d \theta \mathbf{a}_2+d z \mathbf{a}_3
\end{align}
where $(x^1,x^2,x^3)=(r,\theta,z)$. Here, $d \theta$ does not have the same dimensions as the others. The \emph{physical components} in this case are $(dr,rd\theta,dz)$.

The physical components $\tilde{u}^{i}$ of a vector $\mathbf{u}$ are defined to be the components along the \emph{covariant} base vectors (and hence are obtained from the contravariant components), referred to unit vectors. Thus,
\begin{align} % requires amsmath; align* for no eq. number
   \mathbf{u}&=u^i\mathbf{a}_i  \\
                    &=\sum_{i=1}^3 u^ih_i \mathbf{\hat{a}}_i  \\
                    &\equiv \tilde{u}^{i} \mathbf{\hat{a}}_i
\end{align}
and
\begin{align} % requires amsmath; align* for no eq. number
   \tilde{u}^{i} =u^ih_i=u^i\sqrt{G_{ii}}  \quad \textrm{(no summation)}
\end{align}
is called \emph{ physical components of a vector}.

\end{itemize}

\subsection{Fundamental Metric Components}
Using the two types of basis vectors, we can form the scalars
\begin{align} % requires amsmath; align* for no eq. number
   G_{ij}&=\mathbf{a}_i \cdot \mathbf{a}_j=\mathbf{a}_j\cdot \mathbf{a}_i  \\
   G^{ij}&=\mathbf{a}^i \cdot \mathbf{a}^j=\mathbf{a}^j\cdot \mathbf{a}^i
\end{align}
which are the fundamental metric components of the space in which the curvilinear coordinates have been introduced. The components $G_{ij}$ and $G^{ij}$ are the covariant and contravariant components, respectively, of a tensor, called the \emph{metric tensor} and both symmetric. They have a important property
\begin{align} % requires amsmath; align* for no eq. number
   G_{ij}G^{jk} = \delta^k_i,
\end{align}
where $\delta^i_j$ in tensor form is
\begin{align}
 \delta^i_j
 =\begin{bmatrix}
    1 & 0 & 0  \\
    0 & 1 & 0  \\
    0 & 0 & 1  
   \end{bmatrix}.
\end{align}

The following formulas, \emph{ which are of great importance}, can be established by
\begin{align} % requires amsmath; align* for no eq. number
   \mathbf{a}_i & =G_{ij}\mathbf{a}^j    \label{eq:basevectorrelation_1}  \\
   \mathbf{a}^i & =G^{ij}\mathbf{a}_j    \label{eq:basevectorrelation_2} \\
   u_i &  =G_{ik}u^k   \label{eq:vectorrelation_1} \\
    u^j & =G^{jk} u_k   \label{eq:vectorrelation_2}
\end{align}

The determinant of the covariant metric tensor is denoted by
\begin{align} % requires amsmath; align* for no eq. number
   G=\det (G_{ij})
\end{align}
Generally the \emph{Jacobian of transformation} in the curvilinear coordinates is defined as  $\sqrt{G}$.

\subsection{Elemental Displacement Vector}
In a coordinate system $x^i$, the position vector $\mathbf{r}$ and its increment at any point can be written as
\begin{align} % requires amsmath; align* for no eq. number
   d \mathbf{r} &=\frac{\partial \mathbf{r}}{\partial x^i} d x^i   \label{eq:elemDisplacement}  \\
                       &=\mathbf{a}_i dx^i
\end{align}

The magnitude of $d \mathbf{r}$, denoted as $ds$, is defined by
\begin{align} % requires amsmath; align* for no eq. number
   (ds)^2&=d \mathbf{r} \cdot d \mathbf{r}  \\
             &=(\mathbf{a}_i\cdot\mathbf{a}_j)dx^idx^j  \\
             &=G_{ij} dx^idx^j
\end{align}

\begin{itemize}
\item Arc length element 

An increment of arc length on a coordinate line along which $x^i$ varied is given by
\begin{align} % requires amsmath; align* for no eq. number
   ds^i=|\mathbf{a}_i|dx^i
\end{align}

\item  Surface area element

An increment of area on a coordinate surface of constant $x^i$ is given by
\begin{align} % requires amsmath; align* for no eq. number
   dS^i &=|\mathbf{a}_j\times\mathbf{a}_k|dx^jdx^k \\
           &=\sqrt{G_{jj}G_{kk}-G_{jk}^2} dx^jdx^k
\end{align}

\item Volume element
An increment of volume is given by
\begin{align} % requires amsmath; align* for no eq. number
   dV&=\mathbf{a}_1 \cdot (\mathbf{a}_2 \times \mathbf{a}_3)dx^1dx^2dx^3  \\
       &=\sqrt{G}dx^1dx^2dx^3 
\end{align}
\end{itemize}

\subsection{Differentiation of Base Vectors}
The operations of grad, curl, and div on vectors and tensors require a knowledge of partial derivatives of the base vectors in the next. Here some connections among the base vectors are presented.

\begin{itemize}
\item The derivative of covariant base vector

The first one is
\begin{align} % requires amsmath; align* for no eq. number
   \frac{\partial \mathbf{a}_i}{\partial x^j}&=\frac{\partial }{\partial x^j}\left(  \frac{\partial \mathbf{r}}{\partial x^i} \right)=\frac{\partial }{\partial x^i}\left(  \frac{\partial \mathbf{r}}{\partial x^j} \right)   \\
   &=\frac{\partial \mathbf{a}_j}{\partial x^i}
\end{align} 

Based on the definition of metric tensor, i.e., $G_{ij}=\mathbf{a}_i\cdot\mathbf{a}_j$, we differentiate covariant metric tensor and get by mathematical operations
\begin{align} % requires amsmath; align* for no eq. number
   \frac{\partial \mathbf{a}_i}{\partial x^j} \cdot\mathbf{a}_k=[ij,k]
\end{align}
which implies
\begin{align} % requires amsmath; align* for no eq. number
  \frac{\partial \mathbf{a}_i}{\partial x^j}=[ij,k]\mathbf{a}^k  \label{eq:dfbasevec}
\end{align}
where
\begin{align} % requires amsmath; align* for no eq. number
   [ij,k]=\frac{1}{2}\left( \frac{\partial G_{ik}}{\partial x^j}+\frac{\partial G_{jk}}{\partial x^i}-\frac{\partial G_{ij}}{\partial x^k}\right)
\end{align}
are called \emph{Christoffel symbols of the first kind}. 

Both sides of  Eq. \eqref{eq:dfbasevec} are multiplied scalarly by $\mathbf{a}^m$ to get
\begin{align} % requires amsmath; align* for no eq. number
   \frac{\partial \mathbf{a}_i}{\partial x^j}\cdot\mathbf{a}^m=\Gamma_{ij}^m
\end{align}
which implies
\begin{align} % requires amsmath; align* for no eq. number
    \frac{\partial \mathbf{a}_i}{\partial x^j}=\Gamma_{ij}^m\mathbf{a}_m  \label{eq:dfbasevec2}
\end{align}
where
\begin{align} % requires amsmath; align* for no eq. number
   \Gamma_{ij}^m=G^{mk}[ij,k]   \label{eq:2ndcChsym}
\end{align}
are called \emph{Christoffel symbols of the second kind}. Sometimes it is also denoted by
\begin{align} % requires amsmath; align* for no eq. number
  \Gamma_{ij}^m=\left\{ 
  \begin{array}{c}
        m  \\
        i  \quad  j 
   \end{array}
   \right\}.
\end{align}

\item The derivative of contravariant base vector

In a similar way, we obtain the the derivative of contravariant base vector
\begin{align} % requires amsmath; align* for no eq. number
   \frac{\partial \mathbf{a}^i}{\partial x^k}\cdot\mathbf{a}_j=-\Gamma_{jk}^i
\end{align}
which implies
\begin{align} % requires amsmath; align* for no eq. number
   \frac{\partial \mathbf{a}^i}{\partial x^k}=-\Gamma_{rk}^i\mathbf{a}^r  \label{eq:dfbasevec3}
\end{align}

Noted that the relations hold
\begin{align} % requires amsmath; align* for no eq. number
   [ij,k]&=[ji,k]  \\
   \Gamma_{jk}^i&=\Gamma_{kj}^i
\end{align}

\item Useful expressions

When setting $m=i$ in the Eq. \eqref{eq:2ndcChsym} and carrying out summation over the repeated indices, it reaches that
\begin{align} % requires amsmath; align* for no eq. number
   \Gamma_{ij}^i=\frac{1}{2G}\frac{\partial G}{\partial x^j}=\frac{\partial \ln \sqrt{G}}{\partial x^j}. \label{eq:dfbasevec4}
\end{align}
From Eq. \eqref{eq:dfbasevec3} and Eq. \eqref{eq:dfbasevec4}, we have the formula
\begin{align} % requires amsmath; align* for no eq. number
   \frac{\partial}{\partial x^j} \left( \sqrt{G} \mathbf{a}^j \right)=0  \label{eq:dfbasevec5}.
\end{align}
\end{itemize}
This identity is also derived by Thompson et al. (1985), c.f. their Eq. (40).

\subsection{Gradient operator}
In curvilinear coordinates, the grad ($\nabla$) operator is
\begin{align} % requires amsmath; align* for no eq. number
   \textrm {grad} =\frac{\partial}{\partial x^k} \mathbf{a}^k
\end{align}
Note that grad is defined in terms of \emph {covariant components and the contravariant basis}.

\begin{itemize}
\item The gradient of a vector

The gradient of a vector $\mathbf{u}$ read via the definition of gradient operator is
\begin{align} % requires amsmath; align* for no eq. number
   \textrm {grad }  \mathbf{u}=\frac{\partial \mathbf{u}}{\partial x^i} \mathbf{a}^i
\end{align}
\begin{itemize}
\item The covariant derivative of contravariant component of a vector

Substituting $\mathbf{u}=u^k\mathbf{a}_k$ in $\partial \mathbf{u}/\partial x^i$ and using Eq. \eqref{eq:dfbasevec2} (involving the derivative of base vector), we obtain
\begin{align} % requires amsmath; align* for no eq. number
  \frac{\partial \mathbf{u}}{\partial x^i}=u^k_{,i} \mathbf{a}_k  \label{eq:vectorderivative}
\end{align}
where
\begin{align} % requires amsmath; align* for no eq. number
   u^k_{,i}=\frac{\partial u^k}{\partial x^i}+u^r\Gamma_{ir}^k  \label{eq:vectorderivativeExpress}
\end{align}
are called  the \emph{covariant derivative of the contravariant components}.

Thus
\begin{align} % requires amsmath; align* for no eq. number
   \textrm{grads } \mathbf{u}=u^k_{,i}\mathbf{a}_k\mathbf{a}^i
 \end{align}
\end{itemize}

\begin{itemize}
\item The covariant derivative of the covariant components of a vector

Substituting $\mathbf{u}=u_k\mathbf{a}^k$ in $\partial \mathbf{u}/\partial x^i$ and using Eq. \eqref{eq:dfbasevec3}, we get 
\begin{align} % requires amsmath; align* for no eq. number
  \frac{\partial \mathbf{u}}{\partial x^i}&=u_{k,i} \mathbf{a}^k  \label{eq:vectorderivcovariant} \\
  \textrm{grads } \mathbf{u}&=u_{k,i}\mathbf{a}^k\mathbf{a}^i
\end{align}
where
\begin{align} % requires amsmath; align* for no eq. number
   u_{k,i}=\frac{\partial u_k}{\partial x^i}-u_r\Gamma_{ik}^r
\end{align}
is called the \emph{covariant derivative of the covariant components}.

\end{itemize}

\item Gradient of a scalar

For a scalar, the gradient is simply
\begin{align} % requires amsmath; align* for no eq. number\mathbf{a}^i 
  \textrm{grad } \phi =\frac{\partial \phi}{\partial x^i}\mathbf{a}^i =\phi_{,i}\mathbf{a}^i 
\end{align}
by using Eq. \eqref{eq:basevectorrelation_2}, the \emph{nonconservative form} of the gradient reachs
\begin{align} % requires amsmath; align* for no eq. number\mathbf{a}^i 
  \textrm{grad } \phi =\frac{\partial \phi}{\partial x^i}\mathbf{a}^i =G^{ij}\frac{\partial \phi}{\partial x^j}\mathbf{a}_i   \label{eq:gradientnonconservativeform}
\end{align}

Through using Eq. \eqref{eq:dfbasevec5}, the \emph{conservative form} of the gradient is
\begin{align} % requires amsmath; align* for no eq. number\mathbf{a}^i 
  \textrm{grad } \phi =\frac{1}{\sqrt{G}}\frac{\partial (\sqrt{G}\mathbf{a}^j\phi)}{\partial x^j}
\end{align}
using Eq. \eqref{eq:basevectorrelation_2}, it becomes in the covariant base vector
\begin{align} % requires amsmath; align* for no eq. number\mathbf{a}^i 
  \textrm{grad } \phi =\frac{1}{\sqrt{G}}\frac{\partial (\sqrt{G}G^{ij}\mathbf{a}_i\phi)}{\partial x^j}
\end{align}
Furthermore, by partially differentiating the above equation and using Eq. \eqref{eq:dfbasevec2} it reaches
\begin{align} % requires amsmath; align* for no eq. number\mathbf{a}^i 
  \textrm{grad } \phi &=\frac{1}{\sqrt{G}}\frac{\partial (\sqrt{G}G^{ij}\phi)}{\partial x^j}\mathbf{a}_i + G^{ij}\phi\frac{\partial \mathbf{a}_i}{\partial x^j}  \nonumber  \\
  &=\frac{1}{\sqrt{G}}\frac{\partial (\sqrt{G}G^{ij}\phi)}{\partial x^j}\mathbf{a}_i + G^{mn}\phi\Gamma_{mn}^i \mathbf{a}_i \label{eq:scalargradincobasevector}
\end{align}

\end{itemize}

\subsection{Divergence and Curl of a Vector}
\begin{itemize}
\item Divergence of a vector

The divergence of a vector $\mathbf{u}$ is defined by
\begin{align} % requires amsmath; align* for no eq. number\mathbf{a}^i 
  \textrm{div } \mathbf{u} =\nabla \cdot \mathbf{u}=\frac{\partial \mathbf{u}}{\partial x^i}\cdot \mathbf{a}^i.
\end{align}

Based on Eq. \eqref{eq:KroSym} and \eqref{eq:vectorderivative}, we can obtain
\begin{align} % requires amsmath; align* for no eq. number\mathbf{a}^i 
  \textrm{div } \mathbf{u} =u^i_{,i}
\end{align}
where by using Eq. \eqref{eq:dfbasevec4} and Eq. \eqref{eq:vectorderivativeExpress}
\begin{align} % requires amsmath; align* for no eq. number
   u^i_{,i}= \frac{\partial u^i}{\partial x^i}+u^j\frac{\partial }{\partial x^j} \left( \ln \sqrt{G} \right)
\end{align}
Thus,
\begin{align} % requires amsmath; align* for no eq. number\mathbf{a}^i 
  \textrm{div } \mathbf{u} =\frac{1}{\sqrt{G}}\frac{\partial }{\partial x^i} \left( \sqrt{G}u^i \right) =\frac{1}{\sqrt{G}}\frac{\partial }{\partial x^i} \left( \sqrt{G} \mathbf{u}\cdot \mathbf{a}^i \right)
\end{align}

If covariant components of $\mathbf{u}$ are used, then by Eq. \eqref{eq:vectorderivcovariant} we get
\begin{align} % requires amsmath; align* for no eq. number\mathbf{a}^i 
  \textrm{div } \mathbf{u} =G^{ik}u_{i,k}
\end{align}
\end{itemize}

\begin{itemize}

\item Curl of a vector

The definition of Curl of a vector is
\begin{align} % requires amsmath; align* for no eq. number
   \textrm{Curl } \mathbf{u} =\nabla \times \mathbf{u}= \lim_{\Delta V \rightarrow 0} \frac{\int dS \textrm{ }\mathbf{n}\times \mathbf{u}}{\Delta V}
\end{align}
where $V$ is the volume and $\mathbf{n}$ is outward normal unit of the surface $S$. The integrant of curl of a vector over a surface can be defined as
\begin{align} % requires amsmath; align* for no eq. number
   \int dS \textrm{ }\mathbf{n} \cdot \textrm{Curl } \mathbf{u} = \oint_C d \mathbf{r} \cdot \mathbf{u},
\end{align}
where $C$ is the perimeter of the surface $S$. It will be probably used later.  Another important thing is that the curl of gradient dissapear, that is,
\begin{align} % requires amsmath; align* for no eq. number
   \nabla \times \nabla \phi =0
\end{align}
where $\phi$ is a scalar. If $\nabla \times \mathbf{u}=0$, then $\mathbf{u}$ can be written as $\nabla \phi$.

Introducing permutation symbols $e_{ijk}$ and $e^{ijk}$, one can in general write
\begin{align} % requires amsmath; align* for no eq. number
   e_{ijk} \equiv \mathbf{a}_i\cdot (\mathbf{a}_j \times \mathbf{a}_k) &=\epsilon_{ijk} \sqrt{G},  \\
   e^{ijk} \equiv \mathbf{a}^i\cdot (\mathbf{a}^j \times \mathbf{a}^k) &=\epsilon^{ijk} \frac{1}{\sqrt{G} },
\end{align}

then the following identities read
\begin{align} % requires amsmath; align* for no eq. number
   \mathbf{a}_j\times \mathbf{a}_k&=e_{ijk}\mathbf{a}^i=\sqrt{G} \epsilon_{ijk}\mathbf{a}^i  \label{eq:identity1}  \\
   \mathbf{a}^j\times \mathbf{a}^k&=e^{ijk}\mathbf{a}_i=\frac{1}{\sqrt{G}} \epsilon^{ijk} \mathbf{a}_i  \label{eq:identity2} 
\end{align}
where both $\epsilon_{ijk}$ and $\epsilon^{ijk}$ are the permutation symbols, and they have the values
\begin{align} % requires amsmath; align* for no eq. number
   \epsilon_{ijk}= \epsilon^{ijk}=\left \{ \begin{array}{ll}
   1, & \textrm{if } (i,j,k) \textrm{ is } (1,2,3), (3,1,2),\textrm{ or }(2,3,1)\\
   -1, & \textrm{if } (i,j,k) \textrm{ is } (1,3,2), (3,2,1),\textrm{ or }(2,1,3)\\
   0 , & \textrm{otherwise.}
   \end{array} \right.
\end{align}

Noted that the relations hold
\begin{align} % requires amsmath; align* for no eq. number
   e^{ijk}e_{pqr}=\epsilon^{ijk}\epsilon_{pqr},\quad e^{ijk}e_{pqr}= \delta_p^i\delta_q^j-\delta_p^j\delta_q^i
\end{align}

From Eq. \eqref{eq:identity1} and Eq. \eqref{eq:identity2} we deduce that
\begin{align} % requires amsmath; align* for no eq. number
    \mathbf{a}^i&=\frac{1}{2\sqrt{G}}\epsilon^{ijk}(\mathbf{a}_j\times\mathbf{a}_k) \label{eq:identity3}  \\
    \mathbf{a}_i&=\frac{\sqrt{G}}{2}\epsilon_{ijk}(\mathbf{a}^j\times\mathbf{a}^k)  \label{eq:identity4}
\end{align}
thus,
\begin{align} % requires amsmath; align* for no eq. number
   \textrm{Curl } \mathbf{u}=\mathbf{a}^i\times\frac{\partial \mathbf{u}}{\partial x^i}
\end{align}
Using the above developed formulas, we obtain
\begin{align} % requires amsmath; align* for no eq. number
   \textrm{Curl } \mathbf{u}=\frac{1}{\sqrt{G}}\epsilon^{ijk}u_{k,j}\mathbf{a}_i
\end{align}
Thus, the contravariant components of curl $\mathbf{u}$ are
\begin{align} % requires amsmath; align* for no eq. number
   (\textrm{Curl } \mathbf{u})^i=\frac{1}{\sqrt{G}}\left( \frac{\partial u_k}{\partial x^j} - \frac{\partial u_j}{\partial x^k} \right)
\end{align}
where $i,j,k$ are cyclic.

\item The Cross Product of vectors

The cross product of vectors can be written as 
\begin{align} % requires amsmath; align* for no eq. number
   \mathbf{u}\times \mathbf{v} &=e_{kij}u^iv^j \mathbf{a}^k=\sqrt{G} \left | \begin{array}{ccc}
   \mathbf{a}^1 & \mathbf{a}^2 & \mathbf{a}^3  \\
   u^1 & u^2 & u^3 \\
   v^1 & v^2 & v^3
   \end{array} \right|   \\
   &=e^{kij}u_iv_j \mathbf{a}_k=\frac{1}{\sqrt{G}} \left | \begin{array}{ccc}
   \mathbf{a}_1 & \mathbf{a}_2 & \mathbf{a}_3  \\
   u_1 & u_2 & u_3 \\
   v_1 & v_2 & v_3
   \end{array} \right|
\end{align}
where Eq. \eqref{eq:identity1} and Eq. \eqref{eq:identity2} are used.  In another way, we have
\begin{align} % requires amsmath; align* for no eq. number
   \mathbf{u}\times\mathbf{v}&=(u^i\mathbf{a}_i) \times (v^j\mathbf{a}_j)  \\
    &=u^iv^j (\mathbf{a}_i \times \mathbf{a}_j)  \\
    &=u^iv^j e_{kij} \mathbf{a}^k  \quad   \textrm{ using \eqref{eq:identity1}}  \\
    &=u^iv^j e_{kij} G^{kn}\mathbf{a}_n   \quad  \textrm{using \eqref{eq:basevectorrelation_2}}   \\
    &=u^iv^j \sqrt{G}\epsilon_{kij} G^{kn}\mathbf{a}_n
\end{align}
where $n$ is the indices of covariant base vectors and also represents the order of contravariant components. For convenience, we reformulate it as
\begin{align} % requires amsmath; align* for no eq. number
   \mathbf{u}\times\mathbf{v} = u^kv^l \sqrt{G}\epsilon_{jkl} G^{ij}\mathbf{a}_i
\end{align}

\end{itemize}

\subsection{Divergence of Second-Order Tensors}

The divergence of a tensor $\mathbf{T}$ is defined as
\begin{align} % requires amsmath; align* for no eq. number
   \textrm{div } \mathbf{T} =\frac{\partial \mathbf{T}}{\partial x^k}\cdot\mathbf{a}^k
\end{align}

Using the previously defined expressions for the derivatives of base vectors, we have the following
results:

\begin{itemize}
\item Contravariant component of a tensor
\begin{align} % requires amsmath; align* for no eq. number
   \mathbf{T} &=T^{ij}\mathbf{a}_i\mathbf{a}_j  \\
   \textrm{div } \mathbf{T}&=T^{ik}_{,k}\mathbf{a}_i 
\end{align}
where the covariant derivative is
\begin{align} % requires amsmath; align* for no eq. number
   T^{ij}_{,k} &=\frac{\partial T^{ij}}{\partial x^k}+\Gamma_{mk}^iT^{mj}+\Gamma_{mk}^j T^{im} \label{eq:contraderivativeTensor}
\end{align}

Being the contraction of $\Gamma_{mk}^j$ in Eq. \eqref{eq:contraderivativeTensor} by setting $j=k$ 
\begin{align} % requires amsmath; align* for no eq. number
   \Gamma_{mk}^k=\frac{\partial \ln \sqrt{G}}{\partial x^m}=\frac{1}{\sqrt{G}} \frac{\partial \sqrt{G}}{\partial x^m}
\end{align}
the same as Eq. \eqref{eq:dfbasevec4}. Thus,
\begin{align} % requires amsmath; align* for no eq. number
  \textrm{div } \mathbf{T}=T^{ik}_{,k}\mathbf{a}_i &  =\left( \frac{1}{\sqrt{G}} \frac{\partial}{\partial x^k} \left(  \sqrt{G}T^{ik}\right) + \Gamma_{mk}^iT^{mk} \right) \mathbf{a}_i 
\end{align}

\item Covariant component of a tensor
\begin{align} % requires amsmath; align* for no eq. number
   \mathbf{T} &=T_{ij}\mathbf{a}^i\mathbf{a}^j  \\
   \textrm{div } \mathbf{T}&=G^{jk}T_{ij,k}\mathbf{a}_i 
\end{align}
The covariant derivative is
\begin{align} % requires amsmath; align* for no eq. number
   T_{ij,k} =\frac{\partial T_{ij}}{\partial x^k}-\Gamma_{ik}^m T_{mj}-\Gamma_{jk}^m T_{im}
\end{align}

\end{itemize}

%%%%%%%%%%%%%%%%%%%%%%%%%%%%%%%%%%
%%%%%%%%%%%%%   Appendix B   %%%%%%%%%%%%%
%%%%%%%%%%%%%%%%%%%%%%%%%%%%%%%%%%

{\renewcommand\theequation{B\arabic{equation}}%
\setcounter{equation}{0}% reset counter

\section{Geometric summary  in the spherical polar and cubed-sphere coordinates} \label{label:appendixB}

\subsection{The metric tensor}
Consider that the radial base vector is orthogonal to the surface of constant $r$ in the spherical and cubed coordinate and has the unit length, the metric tensor such as $G_{ij}$ or $G^{ij}$ can be decomposed into a 2D component along with a unit radial component
\begin{align} % requires amsmath; align* for no eq. number
   G_{ij}=\left(  
   \begin{array}{cc}
      \overline{G}_{ij}  & 0  \\
      0 & 1
   \end{array}
   \right), \quad 
   G^{ij}=\left(  
   \begin{array}{cc}
      \overline{G}^{ij}  & 0  \\
      0 & 1
   \end{array}
   \right),
\end{align}
where $\overline{G}_{ij}$ and $\overline{G}^{ij}$ are a 2D metric tensor on the constant $r$ surface.

\subsection{Geometrics in the spherical coordinates}
\subsubsection{Base vectors in the spherical polar system}
The covariant base vectors in the spherical polar coordinate are
\begin{align} % requires amsmath; align* for no eq. number
   \mathbf{a}_1&=r \cos \varphi \mathbf{e}_\lambda \label{eq:co_basis1} \\
   \mathbf{a}_2&=r \mathbf{e}_\varphi \label{eq:co_basis2}\\
   \mathbf{a}_3&=\mathbf{e}_r \label{eq:co_basis3}
\end{align}
where $(\mathbf{e}_\lambda,\mathbf{e}_\varphi,\mathbf{e}_r)$ are the local normal unit vectors along the $\lambda$, $\varphi$ and $r$ coordinate direction on sphere. Consider the vector wind $\mathbf{u}=u \mathbf{e}_\lambda + v \mathbf{e}_\varphi+w \mathbf{e}_r$ on sphere, we have 
\begin{align} % requires amsmath; align* for no eq. number
    \mathbf{u}&=u^1\mathbf{a}_1+u^2\mathbf{a}_2+u^3\mathbf{a}_3  \\
    u \mathbf{e}_\lambda + v \mathbf{e}_\varphi+w \mathbf{e}_r&=u^1 r \cos \varphi \mathbf{e}_\lambda + u^2r \mathbf{e}_\varphi+u^3\mathbf{e}_r,
\end{align}
In the matrix form, we get
\begin{align} % requires amsmath; align* for no eq. number
      \begin{pmatrix} % or pmatrix or bmatrix or Bmatrix or ...
         u \\
         v \\
         w \\
      \end{pmatrix}
      =M      \begin{pmatrix} % or pmatrix or bmatrix or Bmatrix or ...
         u^1 \\
         u^2 \\
         u^3 \\
      \end{pmatrix}\label{eq:uvw_relation}
\end{align}
where \begin{align} % requires amsmath; align* for no eq. number
M    = \begin{pmatrix} % or pmatrix or bmatrix or Bmatrix or ...
         r \cos\varphi & 0 & 0\\
         0 & r & 0\\
         0 & 0& 1 \\
      \end{pmatrix} \label{eq:basic_transformation}
\end{align}

Inversing the Eq. \eqref{eq:uvw_relation}, the contravariant velocity components are
\begin{align} % requires amsmath; align* for no eq. number
   (u^1,u^2,u^3)=(u^\lambda,u^\varphi,u^r)=(\frac{u}{r \cos\varphi},\frac{v}{r },w) \label{eq:contra-velo}
\end{align}

\subsubsection{Metrics tensor}
Once we have the spherical transformation matrix \eqref{eq:basic_transformation}, the covariant metric tensor is defined by
\begin{align} % requires amsmath; align* for no eq. number
   G_{ij}&=M^TM \\
            &=\begin{pmatrix} % or pmatrix or bmatrix or Bmatrix or ...
         r^2\cos\varphi^2 & 0 & 0\\
         0 & r^2 & 0\\
         0 & 0& 1 \\
      \end{pmatrix}.
\end{align}
Then the contravariant metric tensor can be attained
\begin{align} % requires amsmath; align* for no eq. number
   G^{ij}&=(G_{ij})^{-1} \\
            &=M^{-1}M^{-T} \\
            &=\begin{pmatrix} % or pmatrix or bmatrix or Bmatrix or ...
        \frac{1} {r^2\cos\varphi^2} & 0 & 0\\
         0 & \frac{1}{r^2} & 0\\
         0 & 0& 1 \\
      \end{pmatrix}.
\end{align}

The Jacobian of the transformation in the spherical polar system is 
\begin{align} % requires amsmath; align* for no eq. number
   \sqrt{G}=\sqrt{\det |G_{ij}|}=\mathbf{a}_1\cdot (\mathbf{a}_2 \times \mathbf{a}_3)=r^2 \cos\varphi.
\end{align}

Noted that $r=R$ ($R$ is the Earth radius) in the \emph{shallow-atmosphere approximation}.

\subsubsection{The Christoffel symbol of the second kind}
From the definition of Eq. \eqref{eq:2ndcChsym}, we obtain the expression of the Christoffel symbols of the second kind in the deep atmosphere
\begin{align} % requires amsmath; align* for no eq. number
   \Gamma^1=   \begin{pmatrix} % or pmatrix or bmatrix or Bmatrix or ...
         0 & -\tan\varphi & \frac{1}{r} \\
         -\tan\varphi & 0 & 0 \\
          \frac{1}{r}  & 0 & 0
      \end{pmatrix}
\end{align}
\begin{align} % requires amsmath; align* for no eq. number
   \Gamma^2=   \begin{pmatrix} % or pmatrix or bmatrix or Bmatrix or ...
         \sin\varphi\cos\varphi & 0 & 0 \\
         0 & 0 & \frac{1}{r}  \\
         0  & \frac{1}{r}  & 0
      \end{pmatrix}
\end{align}
\begin{align} % requires amsmath; align* for no eq. number
   \Gamma^3=   \begin{pmatrix} % or pmatrix or bmatrix or Bmatrix or ...
         -r\cos^2\varphi & 0 & 0 \\
         0 & -r & 0  \\
         0  & 0  & 0
      \end{pmatrix}.
\end{align}

In the shallow-atmosphere approximation, $r$ is constant $r=R$ so that the changed Christoffel symbols take the form
\begin{align} % requires amsmath; align* for no eq. number
   \Gamma^1=   \begin{pmatrix} % or pmatrix or bmatrix or Bmatrix or ...
         0 & -\tan\varphi & 0 \\
         -\tan\varphi & 0 & 0 \\
          0  & 0 & 0
      \end{pmatrix}
\end{align}
\begin{align} % requires amsmath; align* for no eq. number
   \Gamma^2=   \begin{pmatrix} % or pmatrix or bmatrix or Bmatrix or ...
         \sin\varphi\cos\varphi & 0 & 0 \\
         0 & 0 & 0  \\
         0  & 0  & 0
      \end{pmatrix}
\end{align}
\begin{align} % requires amsmath; align* for no eq. number
   \Gamma^3=  0
\end{align}

Noted that the product of Christoffel symbol and contravariant metric  of the spherical polar system in the shallow-atmosphere approximation reads
\begin{align} % requires amsmath; align* for no eq. number
   G^{ij}\Gamma^k_{ij} =\left( \begin{array}{c}
   0 \\
   \tan\varphi/R^2 \\
   0
   \end{array}
   \right),
\end{align}
however, they in the deep-atmosphere approximation are 
\begin{align} % requires amsmath; align* for no eq. number
   G^{ij}\Gamma^k_{ij} =\left( \begin{array}{c}
   0 \\
   \tan\varphi/r^2 \\
   -2/r
   \end{array}
   \right).
\end{align}

\subsection{Geometrics in the cubed-sphere system}
\subsubsection{Base vectors in the cubed-sphere coordinates}
The covariant base vectors in the cubed-sphere coordinate by using Eq. \eqref{eq:elemDisplacement} have the form
\begin{align} % requires amsmath; align* for no eq. number
   \mathbf{a}_1&=\mathbf{r}_\xi=\mathbf{e}_\lambda r \cos \varphi \frac{d \lambda}{d \xi}+ \mathbf{e}_\varphi r \frac{d \varphi}{d \xi} \label{eq:cubed_co_basis1}  \\
   \mathbf{a}_2&=\mathbf{r}_\eta=\mathbf{e}_\lambda r \cos \varphi \frac{d \lambda}{d \eta}+ \mathbf{e}_\varphi r \frac{d \varphi}{d \eta} \label{eq:cubed_co_basis3} \\
   \mathbf{a}_3&=\mathbf{r}_r=\mathbf{e}_r  \label{eq:cubed_co_basis3}
\end{align}
where $(\xi,\eta)=(x^1,x^2)=(\alpha,\beta) \in [-\frac{\pi}{4},\frac{\pi}{4}]\times[-\frac{\pi}{4},\frac{\pi}{4}]$ hold. Consider the wind vector $\mathbf{u}=u\mathbf{e}_\lambda +v\mathbf{e}_\varphi+w\mathbf{e}_r  $ on sphere, the contravariant components of wind vector in the cubed coordinate are related by
\begin{align} % requires amsmath; align* for no eq. number
   u \mathbf{e}_\lambda+v\mathbf{e}_\varphi +w \mathbf{e}_r=\mathbf{u} = u^1 \mathbf{a}_1+u^2 \mathbf{a}_2 +u^3 \mathbf{a}_3 
\end{align}
Put in matrix form
\begin{align} % requires amsmath; align* for no eq. number
           \begin{pmatrix} 
             u  \\
             v  \\
             w \\
            \end{pmatrix}  = {M}  \begin{pmatrix} 
            u^1  \\
             u^2  \\
             u^3  \\
            \end{pmatrix} 
\end{align}
where
\begin{equation} % requires amsmath; align* for no eq. number
   {M} =  
           \begin{pmatrix} 
               r \cos \varphi \lambda_\xi &   r\cos \varphi \lambda_\eta & 0 \\
               r \varphi_\xi                      &    r \varphi_\eta  & 0 \\
               0 & 0&1
            \end{pmatrix} \label{eq:basic_transformation_cubed}
\end{equation}
\subsubsection{Metric tensor}
Once we have the spherical transformation matrix \eqref{eq:basic_transformation_cubed}, the covariant metric tensor is defined by
\begin{align} % requires amsmath; align* for no eq. number
   G_{ij}&=M^TM \\
            &=\begin{pmatrix} % or pmatrix or bmatrix or Bmatrix or ...
            \overline{G}_{ij} &0  \\
            0& 1 \\
      \end{pmatrix}.
\end{align}
where
\begin{align} % requires amsmath; align* for no eq. number
   \overline{G}_{ij}&=\frac{r^2(1+X^2)(1+Y^2)}{\delta^4}\begin{pmatrix} % or pmatrix or bmatrix or Bmatrix or ...
            1+X^2 &-XY \\
            -XY & 1+Y^2 \\
      \end{pmatrix}.
\end{align}
noted that  $X=\tan (x^1)$, $Y=\tan (x^2)$ and $\delta=\sqrt{1+X^2+Y^2}$ are defined.

Then the contravariant metric tensor can be attained
\begin{align} % requires amsmath; align* for no eq. number
   G^{ij}&=(G_{ij})^{-1} \\
            &=M^{-1}M^{-T} \\
            &=\begin{pmatrix} % or pmatrix or bmatrix or Bmatrix or ...
            \overline{G}^{ij} &0   \\
            0& 1 \\
      \end{pmatrix}.
\end{align}
where 
\begin{align} % requires amsmath; align* for no eq. number
   \overline{G}^{ij}&=\frac{\delta^2}{r^2(1+X^2)(1+Y^2)}\begin{pmatrix} % or pmatrix or bmatrix or Bmatrix or ...
            1+Y^2 & XY \\
             XY & 1+X^2 \\
      \end{pmatrix}.
\end{align}

The Jacobian of the transformation in the cubed-sphere coordinates is 
\begin{align} % requires amsmath; align* for no eq. number
   \sqrt{G}=\sqrt{\det |G_{ij}|}=\mathbf{a}_1\cdot (\mathbf{a}_2 \times \mathbf{a}_3)=\frac{r^2(1+X^2)(1+Y^2)}{\delta^3}.
\end{align}

Noted that all $r=R$ ($R$ is the Earth radius) in the \emph{shallow-atmosphere approximation}.

\subsubsection{The Christoffel symbol of the second kind}
From the definition of Eq. \eqref{eq:2ndcChsym}, we write the Christoffel symbols of the second kind in the deep atmosphere as
\begin{align} % requires amsmath; align* for no eq. number
   \Gamma^1=   \begin{pmatrix} % or pmatrix or bmatrix or Bmatrix or ...
         \frac{2XY^2}{\delta^2} & \frac{-Y(1+Y^2)}{\delta^2}& \frac{1}{r} \\
         \frac{-Y(1+Y^2)}{\delta^2} & 0 & 0 \\
          \frac{1}{r}  & 0 & 0
      \end{pmatrix},
\end{align}
\begin{align} % requires amsmath; align* for no eq. number
   \Gamma^2=   \begin{pmatrix} % or pmatrix or bmatrix or Bmatrix or ...
         0 &  \frac{-X(1+X^2)}{\delta^2} & 0 \\
         \frac{-X(1+X^2)}{\delta^2} & \frac{2X^2Y}{\delta^2} & \frac{1}{r}  \\
         0  & \frac{1}{r}  & 0
      \end{pmatrix},
\end{align}
\begin{align} % requires amsmath; align* for no eq. number
   \Gamma^3=  \frac{r(1+X^2)(1+Y^2)}{\delta^4} \begin{pmatrix} % or pmatrix or bmatrix or Bmatrix or ...
         -(1+X^2) & XY & 0 \\
         XY & -(1+Y^2) & 0  \\
         0  & 0  & 0
      \end{pmatrix}.
\end{align}
In the shallow-atmosphere approximation, $r$ becomes constant $R$ and the Christoffel symbol takes the form
\begin{align} % requires amsmath; align* for no eq. number
   \Gamma^1=   \begin{pmatrix} % or pmatrix or bmatrix or Bmatrix or ...
         \frac{2XY^2}{\delta^2} & \frac{-Y(1+Y^2)}{\delta^2}& 0 \\
         \frac{-Y(1+Y^2)}{\delta^2} & 0 & 0 \\
          0  & 0 & 0
      \end{pmatrix},
\end{align}
\begin{align} % requires amsmath; align* for no eq. number
   \Gamma^2=   \begin{pmatrix} % or pmatrix or bmatrix or Bmatrix or ...
         0 &  \frac{-X(1+X^2)}{\delta^2} & 0 \\
         \frac{-X(1+X^2)}{\delta^2} & \frac{2X^2Y}{\delta^2} & 0  \\
         0  & 0  & 0
      \end{pmatrix},
\end{align}
\begin{align} % requires amsmath; align* for no eq. number
   \Gamma^3=  0.
\end{align}

Noted that the product of Christoffel symbol and contravariant metric under the gnomonic mapping in the shallow-atmosphere approximation reads
\begin{align} % requires amsmath; align* for no eq. number
   G^{ij}\Gamma^k_{ij} =0,
\end{align}
however, they in the deep-atmosphere approximation are 
\begin{align} % requires amsmath; align* for no eq. number
   G^{ij}\Gamma^k_{ij} =\left( \begin{array}{c}
   0 \\
   0 \\
   -\frac{2}{r}
   \end{array}
   \right).
\end{align}

\color{black}

\bibliographystyle{elsarticle-num}
\bibliography{<your-bib-database>}



\end{document}